\documentclass[%
 reprint,
groupedaddress,
 amsmath,amssymb,
 aps,
floatfix,
]{revtex4-2}

\usepackage{graphicx}
\usepackage{dcolumn}
\usepackage{bm}
\usepackage{amsmath}
\usepackage{amsthm}
\usepackage{amssymb}
\usepackage{tikz-cd}
\usepackage{braket}
\usepackage{bbm}
\usepackage{graphicx}
\usepackage{lipsum}
\usepackage{float}
\newcommand{\of}[1]{\left( {#1}\right)}

\DeclareMathOperator{\im}{Im}
\DeclareMathOperator{\re}{Re}

\begin{document}

\preprint{APS/123-QED}

\title{Observability of cyclotron resonance in the hydrodynamic regime of bilayer graphene }

\author{Joseph R. Cruise}
 \altaffiliation[]{Department of Physics, Washington University in St. Louis, MO 63104, USA}
\author{Alexander Seidel}
 \altaffiliation[]{Department of Physics, Washington University in St. Louis, MO 63104, USA}
\author{Erik Henriksen}
 \altaffiliation[]{Department of Physics, Washington University in St. Louis, MO 63104, USA}
\author{Giovanni Vignale}
 \altaffiliation[]{The Institute for Functional Intelligent Materials (I-FIM),
 National University of Singapore, 4 Science Drive 2, Singapore 117544}
 \altaffiliation[]{Department of Physics and Astronomy, University of Missouri, Columbia, Missouri 65211, USA}

\date{\today}

\begin{abstract}
We offer theoretical predictions for the frequency of the resonant frequency of transport for the hydrodynamic description of bilayer graphene, as well as provide quantification for the relative strength of this signal throughout phase space. Our calculations are based on classical fluid dynamics equations derived from the Boltzmann equation for bilayer graphene in \cite{simons20}, and suggest that while this resonance is accessible to current experimental techniques, the same mechanism which causes the hydrodynamic resonance to differ from the Fermi liquid value is responsible for a significant broadening of the peak. 

\end{abstract}

\maketitle



\section{\label{pt:1background} Introduction}
Since its isolation in 2004 \cite{novoselov04}, graphene in its various manifestations has demonstrated a remarkable potential for hosting rich phases of quantum matter \cite{crossno_observation_2016,li17,cao18,cao20,Balents2020}. Much of this novel behavior is attributable to the Dirac points in the Brillouin zone, where the excitations behave like (2+1)-d relativistic quasiparticles \cite{novoselov05,Zhang2005}. It has been suggested theoretically  and demonstrated experimentally  that these massless Dirac fermions are realistically described by electron hydrodynamics (EH)\cite{sachdev07,sachdev08b,ryzhii12,succi13,polini13,mirlin15b,narozhny21,koppens12,basov12,polini15,kim16,geim17,falkovich18,ho18} . The recent experimental identification of this novel phase in monolayer graphene has generated interest in other related systems--such as bilayer graphene--for evidence of hydrodynamics \cite{simons20,monch22,tan22}, although this regime has been less thoroughly explored. 

The idea behind EH is tantalizingly simple: charged particles, interacting only amongst themselves through the conservative Coulomb force, will enter a thermal equilibrium, and so on sufficiently long time scales will be described by classical fluid dynamics equations. (See \cite{fong18,narozhny21} for an overview of hydrodynamics in graphene). It is worth emphasizing that this purely classical description is a consequence only of statistical mechanics, and so holds whether the constituent particles are quantum mechanical or not. Hydrodynamics therefore yields analytically tractable classical results even for highly correlated quantum mechanical systems. Unfortunately, the generality of hydrodynamics belies how stringent the requirement of a Coulomb-dominated phase is.

Graphene distinguishes itself among other candidate materials for hydrodynamics by virtue of its stiffness, which reduces electron-phonon scattering, its small Fermi temperature, which mitigates the issue of Pauli blocking, and its compatibility with hexagonal boron nitride, which diminishes the effects of disorder \cite{polini20,dean10}: these characterizations are valid for mono-, as well as bi- and multilayer graphene, all of which are believed to host hydrodynamic phases. (Recent many-body calculations suggest hydrodynamics of AB-stacked bilayer graphene may be even more robust than those of monolayer graphene \cite{tan22}.) The rarity of all these properties occurring simultaneously in a single material is clear from the history of EH, which saw a span of over 40 years between the development of the theory and the first experimental confirmation \cite{gurzhi63}. Consequentially, the fact that different stackings of graphene are all capable of hosting EH demonstrates the importance of this material as a resource for studying this regime. 

In practice, the clearest signatures of hydrodynamics pertain to geometry-dependent viscous effects, such as vortex-induced negative resistances \cite{bandurin16}, parabolic Poiseuille flow \cite{ku20,sulpizio19}, and ratchet effects in spatially modulated samples \cite{ganichev22}. However, hydrodynamics also features unusual microscopic effects which do not depend on boundary conditions, such as a cyclotron resonance which differs from the Fermi liquid prediction of $\omega_c = eB/m^*$; for the relativistic hydrodynamics of monolayer graphene, it has been shown that $\omega_c \sim 1/T$ \cite{sachdev08a}. This non-trivial temperature dependence is made even more striking in recalling Kohn's theorem, which implies that the free-electron value of $eB/m^*$ holds across all single-band, massive, Galilean-invariant quasiparticle excitations \cite{kohn61}. 

In this paper, we will explore the consequences of hydrodynamics on the cyclotron resonance in AB stacked bilayer graphene. In particular, we will compute the longitudinal conductivity $\sigma_{xx}$ in order to search for a region of phase space where the resonance is substantially removed from the typical value of $eB/m^*$. Our starting point is the hydrodynamic equations of motion given by Ngyuen et. al. ins \cite{simons20Comp}. In section II we compute the resonance analytically, and discuss the regimes where our calculation is valid; in section III we plot the resulting thermoelectric transport coefficients, giving clear experimental predictions; we conclude in section IV.

\section{\label{pt1:analytic-results}Analytic results}
\begin{figure}[!htb]\centering
 \includegraphics[width = \linewidth]{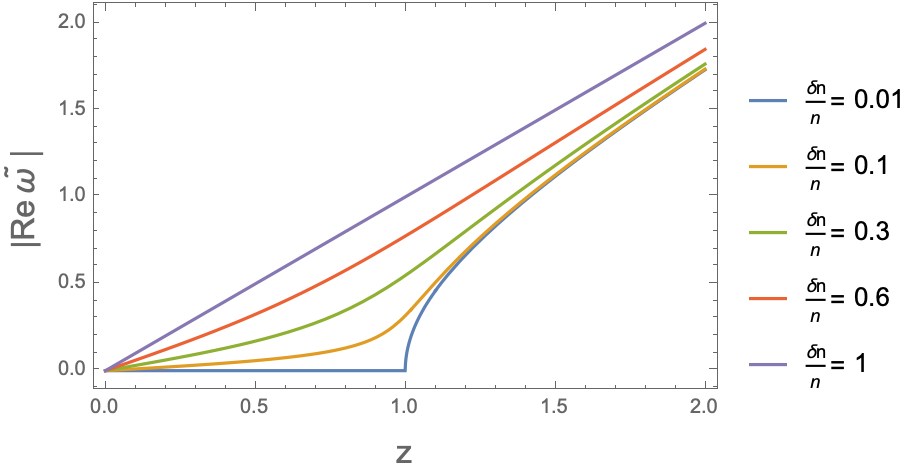}
 \caption{Positive real part of the cyclotron resonance frequency, measured in units of $2\tau_0$. Curves for negative $\delta n/n$ are omitted, as they are identical to positive curves of the same magnitude in $\delta n/n$. } 
 \label{fig:cyResRePart}
 \end{figure}

Following \cite{simons20Comp}, we use the quantum Boltzmann equation (QBE) formalism to derive fluid dynamics equations for both the electrons and holes in bilayer graphene. These equations read
\begin{widetext}
\begin{align}
    -i \omega \vec{u}^e &= - \frac{n^h}{n}\frac{1}{\tau_0} \of{\vec{u}^e - \vec{u}^h} - \frac{\vec{u}^e}{\tau_{d}} + \frac{e}{m_e^*} \of{ \vec{E} + \vec{u}^e \times \vec{B}} - \frac{\Lambda^e}{m^*} k_B \nabla T \label{e} \\
    -i \omega \vec{u}^h &=  \frac{n^e}{n}\frac{1}{\tau_{0}} \of{\vec{u}^e - \vec{u}^h} - \frac{\vec{u}^h}{\tau_{d}} - \frac{e}{m^*} \of{ \vec{E} + \vec{u}^h \times \vec{B}} - \frac{\Lambda^h}{m_h^*} k_B \nabla T. \label{h}
\end{align}
\end{widetext}
Here $\vec{u}^{(e/h)}$, $n^{(e/h)}$, and $\Lambda^{(e/h)}$ are the fluid velocity, number density, and per-particle entropy of the electron/hole fluid respectively, and $n = n^e + n^h$; $\tau_d$ is the disorder scattering time (which accounts for phonon, impurity, and finite size interactions); $\tau_{0}$ is the Coulomb scattering time; and $m^*$ is the effective mass of electrons/holes \footnote{at low energy and magnetic fields, bilayer graphene is particle-hole symmetric, and so for simplicity we will assume both electrons and holes have the same effective mass}. As equations (\ref{e}) and (\ref{h}) are derived in the linear response regime, they can be interpreted as a constitutive relation of the form
\begin{equation}\label{consteq}
    \begin{pmatrix}
    \vec{J} \\ \vec{Q} 
    \end{pmatrix}
    = 
    \begin{pmatrix}
    \sigma & \Theta \\ 
    T\Theta &  K \\
    \end{pmatrix}
    \begin{pmatrix}
    e\vec{E}\\
    k_B \vec\nabla T 
    \end{pmatrix},
\end{equation}
for $\vec{J} = e(n^e \vec{u}^e - n^h \vec{u^h})$ and $Q = k_BT(\Lambda^e n^e \vec{u}^e + \Lambda^h n^h \vec{u}^h) - (\mu/e)\vec{J}$ \footnote{Heat currents are usually measured under the condition $\vec{J} = 0$, in which case $\vec{Q}= -\kappa \vec \nabla T$ where $\kappa := K - T\Theta \sigma^{-1} \Theta$ is the true quantity being measured}. It follows that the longitudinal thermo-electric coefficients are given by 
\begin{widetext}
\begin{align}
    \sigma_{xx}\of{\tilde\omega} &= \sigma_D \frac{ \tilde \omega \of{\of{\frac{\delta n}{n}}^2\of{1+ \tilde \omega} + q(\tilde \omega)}}{\of{\frac{\delta n}{n} \omega_c \tau_0}^2 + q(\tilde \omega)^2}  \label{longcond}\\
    \Theta_{xx}(\tilde \omega) &= -\sigma_D \frac{\tilde \omega \of{\delta \Lambda\of{ \of{\frac{\delta n}{n}}^2 (1+\tilde \omega) + q(\tilde \omega)}+ \Lambda \frac{\delta n}{n}\of{\of{\tilde \omega +1}^2 + \of{\omega_c \tau_0}^2}}}{\of{\frac{\delta n}{n} \omega_c \tau_0}^2 + q(\tilde \omega)^2} \label{longthermpwr}\\
    K_{xx}(\tilde \omega) &= - \kappa_D \left[\frac{\tilde \omega \of{ \of{ \delta \Lambda^2 -\delta \Lambda \beta \mu }\of{ \of{\frac{\delta n}{n}}^2(1+\tilde \omega)+ q(\tilde\omega)} + \Lambda \frac{\delta n}{n} \of{2 \delta \Lambda - \Lambda \beta \mu}\of{ (\tilde \omega+1)^2 + (\tau_0 \omega_c)^2}} }{\of{\frac{\delta n}{n} \omega_c \tau_0}^2 + q(\tilde \omega)^2}\right. \label{longpelt}\\
    & \ \ + \left.\frac{\Lambda^2 \of{ (1+\tilde \omega)q(\tilde \omega)- \of{\frac{\delta n}{n} \tau_0 \omega_c}^2}}{\of{\frac{\delta n}{n} \omega_c \tau_0}^2 + q(\tilde \omega)^2}\right] \nonumber
\end{align}
\end{widetext}
for
\begin{equation}
    q(\omega):= \omega^2 + \omega + (\tau_0 \omega_c)^2.
\end{equation}
Here, we employ the notation: $\sigma_D = ne\tau_0/m$ and $\kappa_D = n k_B^2T \tau_0/m$ are the Drude model electric and thermal conductivities respectively; $\delta n = n^e - n^h$, $ \Lambda = (\Lambda^e+\Lambda^h)/2$ and $\delta \Lambda = (\Lambda^e - \Lambda^h)/2$; $\tilde \omega = -i\omega\tau_0 + \tau_0/\tau_d$. In general, the field coefficients defined in equation (\ref{consteq}) are complex functions, whose real part gives the response function of the material: in this way, we can glean information about their resonant behavior by treating them as rational functions of the complex variable $\tilde \omega$.
Poles of the thermoelectric coefficients, known as cyclotron resonances, correspond to values of $\omega$ for which equations (\ref{e}) and (\ref{h}) are degenerate, i.e. non-invertable. This condition is equivalent to the vanishing of a determinant, and means each response function has the same pole structure. As such, the resonances are the four solutions to the equation
\begin{equation}
    q(\omega_p)^2 = -\of{ \frac{\delta n}{n} \omega_c \tau_0}^2,
\end{equation}
which are
\begin{equation} \label{eqnsix} \omega_p =  - \frac{i}{\tau_d} - \frac{i}{2\tau_0} \pm_2 i\sqrt{\frac{1}{4\tau_0^2} - \omega_c^2 \mp_1 i \frac{\delta n}{n} \frac{\omega_c}{\tau_0}}, \end{equation}
with signs $\pm_1$ and $\pm_2$ uncorrelated. So long as the pole frequency is close to the real axis in the complex plane, the location of the maxima of the thermo-electric coefficients will be well approximated by $\re[\omega_p].$  We are therefore led to study the real and imaginary parts of $\omega_p$, which are
\begin{align}\label{cyresrepart}
    2\tau_0 \re[\omega_p]=
    \begin{cases}
        \displaystyle \pm_{1-2}A\sin \of{ \frac{1}{2} \tan^{-1} x} & 1 > z \\
        \\
        \displaystyle \pm_2 A \cos\of{\frac{1}{2} \tan^{-1} x} & 1 < z,
    \end{cases}
\end{align}
and
\begin{align}\label{cyresimpart}
    2&\tau_0 \im[\omega_p] =  \nonumber\\
    & 
    \begin{cases}
       \displaystyle-(1+2\Gamma_d) \mp_{2}A \cos \of{ \frac{1}{2} \tan^{-1} x} & 1 > z \\
        \\
        \displaystyle-(1+2\Gamma_d) \pm_{2-1} A \sin\of{\frac{1}{2} \tan^{-1} x} & 1 < z,
    \end{cases}
\end{align}
for $\pm_{1-2} = (\pm_1 )(\mp_2 )$, $z:= 2\tau_0 \omega_c$, $x:= (2z\delta n/n)(1-z^2)$, $A = \of{ (1-z^2)^2 + \of{2 (\delta n/n) z}^2}^{1/4}$, and $\Gamma_d := \tau_0/\tau_d$. Notice that the ambiguous sign in front of each expression implies that cyclotron resonances come in pairs: if $\omega$ is a resonant frequency, then $-\omega$ is as well. Experimentally, this means the behavior of the maxima of any thermoelectric conductivity experiment will be determined by the continuous function $|\re[\omega_p]|$, as the temperature, chemical potential, and external magnetic field are varied; see Fig. \ref{fig:cyResRePart} for a plot of this function across both regimes. We remark that $\delta n/n = 1$ describes the case in which there are only electrons and no holes, in whci case the cyclotron frequency reduces to the ordinary Fermi liquid result $\omega_p = eB/m^*$. Equation (\ref{cyresrepart}) tells us the approximate location of the peak of the hydrodynamic cyclotron resonance,  but it does not give us information about the width of that peak, or the accuracy of the approximation; if we want our experimental signals to be sharp, we must look for resonances which have a small imaginary part, i.e. for which $\im[\omega_p]/\re[\omega_p] \ll 1$. 

\begin{figure}[!htb]\centering
 \includegraphics[width=1\linewidth]{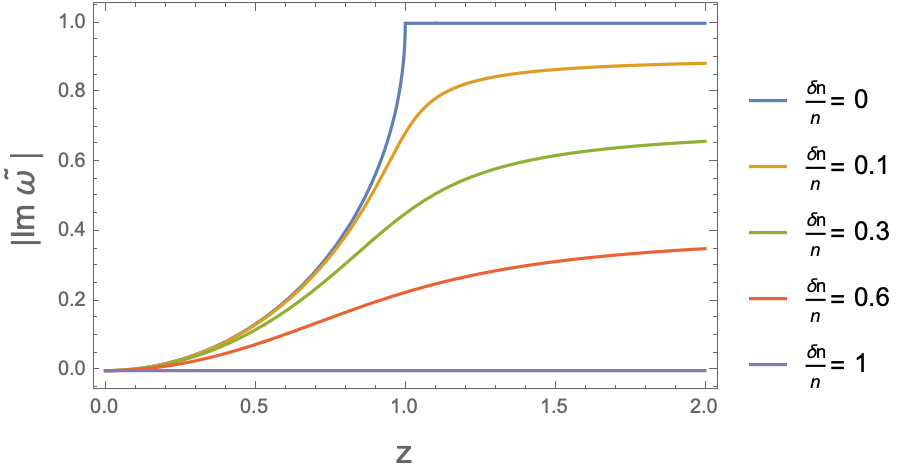}
 \caption{Distance from the real axis of the closest hydrodynamic cyclotron resonance with positive real part, measured in units of $2\tau_0$.} 
 \label{fig:cyResImPart}
 \end{figure}

Unlike for $\re[\omega_p]$, the magnitude of $\im [\omega_p]$ depends on our choice of the signs $\pm_1$ and $\pm_2$.  Assuming that the resonance behavior is dominated by the closest pole to the real axis with positive real part, we are free to take $|\im[\omega_p]|$ to mean the smallest (in magnitude) positive value of the right-hand side of (\ref{cyresimpart}). A plot of this function for $\Gamma_d = 0$, i.e. in the perfectly clean limit, is included in Figure (\ref{fig:cyResImPart}).

\begin{figure}[!htb]\centering
 \includegraphics[width=.9\linewidth]{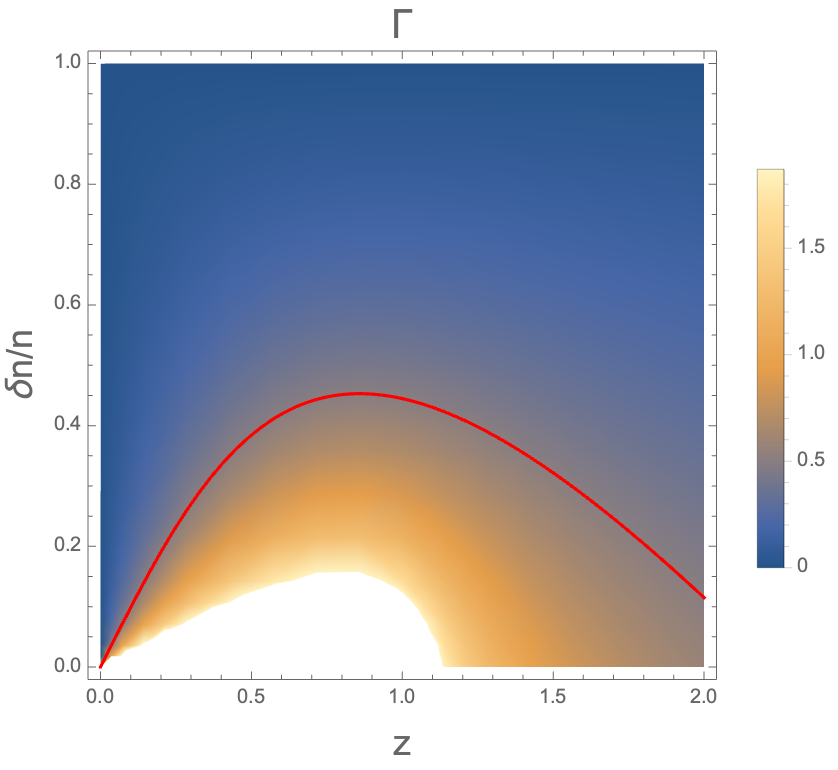}
 \includegraphics[width=.9\linewidth]{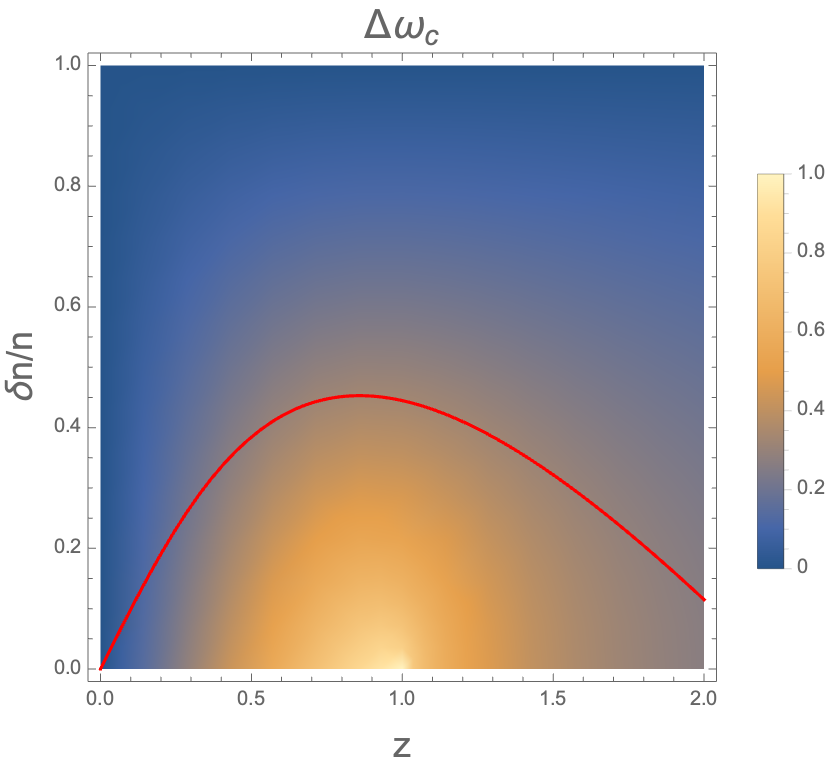}
 \caption{Heat maps for the quality function $\Gamma$ (above), and the difference between the hydrodynamic cyclotron resonance and the Fermi liquid value $\Delta \omega_c$ (below), both as functions of $z$ and $\delta n/n$. The red curve on each plot indicates the $\Gamma = 0.5$ level-set along which $\Delta \omega_c$ is maximized, as discussed in the text.} 
 \label{fig:qualityHeatMap}
 \end{figure}

Comparing Figures \ref{fig:cyResRePart} and \ref{fig:cyResImPart} qualitatively, we see that near $\omega_c\tau_0 = 1/2$, the resonant frequency is maximally displaced from the Fermi liquid result of $\omega_p = \omega_c$ for small $\delta n/n$, but that this region also corresponds to rapid enhancement in the magnitude of $|\im[\omega_p]|$. To gauge the quality of the resonance peak at given values of $z$ and $\delta n/n$, we introduce a function $\Gamma:= |\im[\omega_p]|/|\re[\omega_p]|$. Broadly speaking, $\Gamma$ measures the width of the resonance, normalized by the magnitude of its frequency: an ideal resonance corresponds to $\Gamma = 0$, while $\Gamma \gg 1$  implies a response function which is not determined by resonant behavior. At $\delta n/n = 0.1$, we have $\Gamma:= |\im[\omega_p]|/|\re[\omega_p]| \approx 2.16$, corresponding to a broad peak. A heat map of the quality function $\Gamma = \Gamma(z,\delta n/n)$ is presented in Figure \ref{fig:qualityHeatMap}(a): notably, the signal of a hydrodynamic cyclotron resonance is weakest in exactly the region where the effect of hydrodynamics on the resonance is strongest, i.e. where $\Delta \omega_c := | \re[\omega_p] - \omega_{c}|$ is maximized, as can be seen from Figure \ref{fig:qualityHeatMap}(b). In order to determine the optimal region of phase space for observing hydrodynamics, we will maximize $\Delta \omega_c$, subject to a constraint placed on $\Gamma$, as we will explore numerically in the following section.

In addition to clean system limit, we can also consider the fully compensated, i.e. $\delta n/n$ limit, which occurs at zero chemical potential, and corresponds to an equal density of electrons and holes. For such systems, the equations for the thermo-electric coefficients simplify greatly: for example, the longitudinal conductivity becomes
\begin{equation}\label{fullCompLongCond}
    \sigma_{xx}(\tilde \omega) = \sigma_D \frac{\tilde \omega}{q(\tilde \omega)}.  
\end{equation}
Recalling that the conductivity measured in the laboratory is the real part of equation (\ref{fullCompLongCond}), we find that the frequency which maximizes the conductivity is given by
\begin{align}\label{fullCompRes}
    \of{\omega_{\sigma}^{xx}}^2 &= \frac{ (1+2\Gamma_d) \omega_c\sqrt{q\of{\Gamma_d}}}{(1+ \Gamma_d )\tau_0} -  \frac{\Gamma_dq\of{\Gamma_d}}{(1+ \Gamma_d )\tau_0^2},
\end{align}
and that this corresponds to a maximal value of 
\begin{equation}
    \frac{\of{\sigma_{xx}}_\text{max}}{\sigma_D}=\frac{1 + \Gamma_d - 2 \tau_0\omega_c\of{ \tau_0 \omega_c + \sqrt{q\of{\Gamma_d}}}}{(1+2\Gamma_d)\of{1-(2\tau_0\omega_c)^2}}.
\end{equation}
Since $\Gamma_d$ and $\tau_0 \omega_c$ are always positive, this equation is bounded above by 1, and achieves this value if and only if $\Gamma_d = 0$. $\sigma_D$ therefore represents the ideal limit for hydrodynamic conductivity at charge neutrality in bilayer graphene. 

To conclude this section, we remark that similar calculations can be preformed for the other thermo-electric coefficients, but that at zero chemical potential both particles and holes carry the same entropy so that thermal and electric transport look quite similar. 
\section{\label{pt1:plots}Numerical computation}
\begin{figure}\centering
    \vspace{4 mm}
    \includegraphics[width=\linewidth]{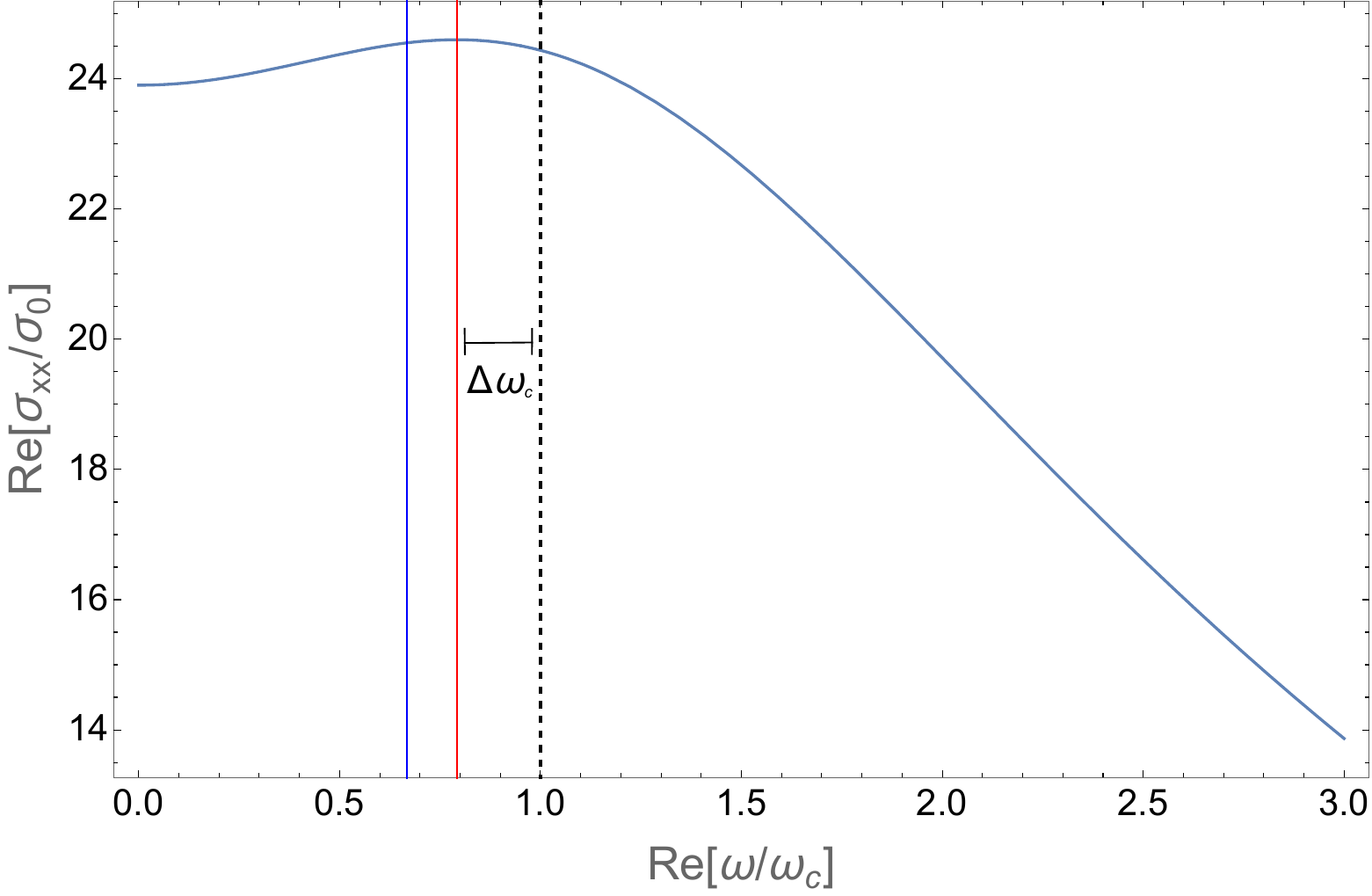}
    \caption{Longitudinal electrical conductivity, $\sigma_{xx}$, plotted in units of $\sigma_0 = e^2/h$. A value of $T = 90$K, $\mu = 0.66/\beta$, and $B = 0.18$T, were chosen for the temperature, chemical potential, and magnetic field respectively. The dashed vertical line represents the classical Fermi liquid cyclotron resonance $\omega_c = eB/m*$, while the red and blue lines are the numerically and theoretically computed values of the hydrodynamic cyclotron resonance $\omega_{p}$ respectively. } 
    \label{fig:longCondz=.999897dn=0.444768}
 \end{figure}

We consider the case of a symmetric two-fluid plasma with $m_e^* = m_h^* = 0.033\cdot m_e$, and for experimentally relevant temperature $T = 90$ K. Values of $B = 0.18$T, and $\mu = 0.66/\beta$ were chosen to maximize $\Delta \omega_c$ subject to the condition $\Gamma = 0.5$---see Figure (\ref{fig:qualityHeatMap}). Model parameters, such as $\tau_0$ and $\tau_d$, are computed in accordance with the supplementary material of \cite{simons20}; using equation (\ref{cyresrepart}), we predict a hydrodynamic cyclotron frequency of $\re[\omega_p] = 0.67 \cdot \omega_c = 0.59$ THz. Figure \ref{fig:longCondz=.999897dn=0.444768} displays the longitudinal electrical conductivity, $\sigma_{xx}$, of the symmetric two-fluid system as a function of the drive frequency $\omega$: the exact maximum in the conductivity occurs at $\omega_\text{max} = 0.79 \cdot \omega_c = 0.70$ THz, corresponding to a displacement $\Delta \omega_c = 0.21 \cdot \omega_c = 1.84 \times 10^{11}$ Hz from the conventional cyclotron resonance. We likewise generated plots for oher thermoelectric coefficients, such as the Hall resistance and the Seebeck coefficient, (\ref{consteq}) using the same parameters as above, however their resonance peaks, being principally determined by the pole structure of $\tilde \omega$, demonstrate no new behaviour, and are not included here. 

\section{Discussion}
\label{pt1:discussion}
We begin this section with a disclaimer: the results given above pertain to homogeneous response in the linear-response regime, and so do not strictly speaking depend on the hydrodynamics of the underlying equations of motion. In fact, it is not clear a priori that the hydrodynamic assumption made in \cite{simons20}, i.e. that the electron and hole fluids equilibriate independently before interacting, holds: however, the authors support their assumption with convincing numerical data, and so we are confident that results derived from these hydrodynamic equations will be valid, if only approximate.  

Generically, non-interacting electron fluids in the presence of disorder have transport properties given, up to constants, by the Drude model\cite{Ashcroft76}. In particular, Kohn's theorem ensures that in isolation, such fluids must have a resonance at exactly $\omega_c$. The physics becomes richer if a second band is introduced, as long as the two bands are allowed to interact, e.g. via the coulomb force. Under such conditions, each fluid drags on the other, potentially leading to a retardation in the resonant frequency from the free electron value of $\omega_c$. However, the same drag force which causes a deviation in resonance frequency is also responsible for the broadening of the signal; this effect is reflected in figures \ref{fig:cyResRePart} and \ref{fig:cyResImPart}, which show that the the region where $\omega_c$ is smaller than $\tau_0^{-1}$ but non-zero corresponds to non-linear dependence of the real part of resonance on the magnetic field, as well as rapid growth in the imaginary part of the resonance. 

Figures \ref{fig:cyResRePart} and \ref{fig:cyResImPart} also reveal the importance of charge compensation to the hydrodynamic effects on the cyclotron resonance. Intuitively, this result is a reflection of a broader principle: whenever a single time scale--coulomb, disorder, magnetic, or anything else--becomes by far the shortest scale in the problem, the picture of interacting fluids breaks down, and we recover traditional Drude transport. Consequently, when the system is far from charge compensation, the drag term of the dominant fluid will be suppressed by a factor of $n^{(e/h)}/n$, meaning the non-linearity of the resonance will vanish, as we see in figure \ref{fig:cyResRePart}; this corresponds to the case of a single fluid interacting weakly with a background through the coulomb force. 

Finally, we can see this principle at work in equation (\ref{fullCompLongCond}): as mentioned in that section, $\sigma_{xx}$ is bounded above by $\sigma_D= ne\tau_0/m$, which is the Drude conductivity except with the usual disorder timescale $\tau_d$ replaced by the coulomb drag $\tau_0$, and that this limit is achieved exactly when $\Gamma_d = 0$. Furthermore its straightforward to check that $\omega_\sigma^{xx}$, defined in equation (\ref{fullCompRes}), reduces to $\omega_c$ in the limit $\Gamma_d \rightarrow 0$, as we would expect. We then see that the unusual behavior of the hydrodynamic resonance in two-fluid hydrodynamic bilayer graphene is a consequence of a competition between interaction time scales, and which interpolates between limits of Drude transport whenever one single time scale is allowed to dominate.

\section{\label{pt1:conclusion}Conclusion}
In this paper, we found analytic formulae for the thermoelectric coefficients in the two-fluid hydrodynamic regime of bilayer graphene, starting from the fluid dynamic equations \eqref{e} and \eqref{h}. Our results indicate a departure from the traditional cyclotron frequency $eB/m^*$ predicted by Kohn's theorem, revealing instead a characteristic dependence on the magnetic field and carrier densities in the hydrodynamic regime of bilayer graphene. Our work also suggests target conditions for the direct observation of a hydrodynamic resonance in real bilayer graphene systems.

\begin{acknowledgments}
 JRC and AS gratefully acknowledge support under NSF Grant No. DMR-2029401.
EH acknowledges support under NSF CAREER DMR-1945278. G.V. was supported by the Ministry of Education, Singapore, under its Research Centre of Excellence award to the Institute for Functional Intelligent Materials (I-FIM, project No. EDUNC-33-18-279-V12)
\end{acknowledgments}

\bibliography{main}

\end{document}